\documentclass[useAMS,usenatbib]{mn2e}

\newcommand{\beq}{\begin{equation}}
\newcommand{\eeq}{\end{equation}}
\newcommand{\bdm}{\begin{displaymath}}
\newcommand{\edm}{\end{displaymath}}
\newcommand{\bea}{\begin{eqnarray}}
\newcommand{\eea}{\end{eqnarray}}
\newcommand{\bt}{\begin{tabular}}
\newcommand{\et}{\end{tabular}}

\usepackage{graphicx}
\usepackage{times}
\usepackage{psfig,epsfig} 
\usepackage{natbib}

\newcommand{\fnl}{f_{\rm{NL}}}

\def\PRL{Phys. Rev. Lett.}
\def\JCAP{J. Cosmol.  Astropart. Phys.}
\def\aap{A\&A}
\def\apj{ApJ}

\def\mnras{MNRAS}

\def\NAR{New Astron. Rev.}

\def\physrep{Phys. Rep.}
\def\nat{Nat}

\def\apjs{ApJS}
\def\prd{Phys. Rev. D}

\title[Haloes in non-Gaussian Models]
{Evolution of Massive Haloes in non-Gaussian Scenarios}
\author[M. Grossi et al.] 
{M. Grossi$^{1,2}$, 
K. Dolag$^{1}$,
E. Branchini$^{3}$,
S. Matarrese$^{4,5}$, 
L.Moscardini$^{2,6}$\\
$^1$ Max-Planck Institut fuer Astrophysik,
Karl-Schwarzschild Strasse 1, D-85748 Garching, Germany
(margot,kdolag@mpa-garching.mpg.de)\\
$^2$ Dipartimento di Astronomia, Universit\`a di Bologna,
via Ranzani 1, I-40127 Bologna, Italy 
(lauro.moscardini@unibo.it)\\
$^{3}$ Dipartimento di Fisica, Universit\`a di Roma TRE,
via della Vasca Navale 84, I-00146, Roma, Italy
(branchin@fis.uniroma3.it)\\
$^{4}$ Dipartimento di Fisica, Universit\`a di Padova,
via Marzolo 8, I-35131, Padova, Italy
(sabino.matarrese@pd.infn.it)\\
$^{5}$ INFN, Sezione di Padova,
via Marzolo 8, I-35131, Padova, Italy\\
$^{6}$ INFN, Sezione di Bologna, viale Berti Pichat 6/2,
I-40127 Bologna, Italy\\
}

\date{}
\begin{document}

\date{Accepted ???. Received ???; in original form July 2007}

\pagerange{\pageref{firstpage}--\pageref{lastpage}} \pubyear{2007}

\maketitle

\label{firstpage}

\begin{abstract}
We have performed high-resolution cosmological N-body simulations of a
{\it concordance} $\Lambda$CDM model to study the evolution of
virialized, dark matter haloes in the presence of primordial
non-Gaussianity. Following a standard procedure, departures from
Gaussianity are modeled through a quadratic Gaussian term in the
primordial gravitational potential, characterized by a dimensionless
non-linearity strength parameter $f_{\rm NL}$.  We find that the halo
mass function and its redshift evolution closely follow the analytic
predictions of \citep{matarrese2000}. The existence of precise
analytic predictions makes the observation of rare, massive objects at
large redshift an even more attractive test to detect primordial
non-Gaussian features in the large scale structure of the universe.
\end{abstract}



\section{Introduction} \label{sect:intro}

Inflation is considered the dominant paradigm for understanding the
initial conditions for structure formation in the Universe.  As a
consequence of the assumed flatness of the inflaton potential, any
intrinsic non-linear (hence non-Gaussian, NG) effect during standard
single-field slow-roll inflation is generally small. Thus, adiabatic
perturbations originated by the quantum fluctuations of the inflaton
field during standard inflation are nearly Gaussian distributed.
Despite the simplicity of the inflationary paradigm, however, the
mechanism by which perturbations are generated is not yet fully
established and various alternatives to the standard scenario have
been considered. A key-point is that the primordial NG is
model-dependent. While standard single-field models of slow-roll
inflation lead to small departures from Gaussianity, non-standard
scenarios for the generation of primordial perturbations in
single-field or multi-field inflation allow for a larger level of
non-Gaussianity \citep[see, e.g.][]{bartolo2002, bernardeau2002,
chen2007}. Moreover, alternative scenarios for the generation of the
cosmological perturbations like the curvaton
\citep{lyth2003, bartolo2004b}, the inhomogeneous reheating
\citep{kofman2003,dvali2004} and the Dirac-Born-Infeld(DBI)-inflation
\citep{alishahiha2004} scenarios, are characterized by a potentially
large NG level \cite[see for a review][]{bartolo2004}.

Because it directly probes the primordial perturbation field, the
Cosmic Microwave Background (CMB) temperature anisotropy and
polarization pattern has been considered the preferential way for
detecting, or constraining, primordial NG signals, thereby shedding
light on the physical mechanisms for perturbation generation.

Alternatively, one can consider the Large-Scale Structure (LSS) of the
Universe. This approach has both advantages and disadvantages.  Unlike
the CMB, which refers to a 2D dataset, LSS carries information on the
3D primordial fluctuation fields. On the other hand, the late
non-linear evolution introduces NG features on its own, that need to
be disentangled from the primordial ones. This can be done in two
different ways: observing the high-redshift universe, such as e.g.
anisotropies in the 21cm background \citep[][see also
\cite{cooray2006}]{pillepich2007}, or studying the statistics of rare
events, such as massive clusters, which are sensitive to the tails of
the distribution of primordial fluctuations.

So far, the investigation of the effects of NG models in the LSS 
has been carried out primarily by analytic means
\citep{lucchin1988,koyama1999,matarrese2000,robinson2000a,robinson2000b,verde2000,
verde2001,komatsu2003,scoccimarro2004,amara2004,ribeiro2007,sefusatti2007,
sefusatti2007b, sadeh2007}. N-body simulations with NG initial
conditions were performed in the early nineties
\citep{messina1990,moscardini1991,weinberg1992}, but the considered
NG models were rather simplistic, being based on fairly
arbitrary distributions for the primordial density fluctuations
\citep[see also][]{mathis2004}. Only very recently, \cite{kang2007}
have investigated more realistic NG models.

The suite of new N-body experiments presented in this paper considers,
for the first time, large cosmological volumes at very high-resolution
for physically motivated NG models (as described in the following
section).  Here we focus on massive haloes and their redshift
evolution.  A more general and exhaustive presentation of the
simulations and the main results will be done elsewhere (Grossi et
al., in preparation).

This paper is organized as follows. In Section~\ref{sect:nong} we
introduce the NG model here considered. The characteristics of our
N-body simulations are presented in Section~\ref{sect:simul}.  In
Section~\ref{sect:halomass} we compute the halo mass function and
compare it with analytic predictions.  We discuss the results and
conclude in Section~\ref{sect:discussion}.

\section{Non-Gaussian Models} \label{sect:nong}

For a large class of models for the generation of the initial seeds
for structure formation, including standard single-field and
multi-field inflation, the curvaton and the inhomogeneous reheating
scenarios, the level of primordial non-Gaussianity can be modeled
through a quadratic term in the Bardeen's gauge-invariant
potential\footnote{on scales much smaller than the Hubble radius,
Bardeen's gauge-invariant potential reduces to minus the usual
peculiar gravitational potential} $\Phi$, namely
\begin{equation}
\label{FNL}
\Phi = \Phi_{\rm L} + f_{\rm{NL}} \left(\Phi_{\rm L}^2 - 
\langle\Phi_{\rm L}^2\rangle \right) \;,
\end{equation}
where $\Phi_{\rm L}$ is a Gaussian random field and the specific value
of the dimensionless non-linearity parameter $f_{\rm{NL}}$ depends on
the assumed scenario \citep[see, e.g.,][]{bartolo2004}.

It is worth stressing that eq.(\ref{FNL}), even though commonly used,
is not generally valid: detailed second-order calculations of the
evolution of perturbations from the inflationary period to the present
time show that the quadratic, non-Gaussian contribution to the
gravitational potential should be represented as a convolution with a
kernel $f_{\rm NL}({\bf x},{\bf y})$ rather than a product
\citep{bartolo2005}.  However, for $|f_{\rm NL}| \gg 1$ as assumed in
this paper, all space-dependent contributions to $f_{\rm NL}$ can be
neglected and the non-linearity parameter can be effectively
approximated by a constant.  In this work we will take $-1000 \le
f_{\rm NL} \le +1000$. We notice that owing to the smallness of $\Phi$,
the contribution of non-Gaussianity implied by these values of $f_{\rm
NL}$ is always within the percent level of the total primordial
gravitational potential, and does not appreciably affect the linear
matter power spectrum.

Yet, this range is larger than that of $-54 < f_{\rm NL} < 114$,
currently allowed, at the 95 per cent confidence level, by CMB data
\citep{spergel2007}.  The rationale behind this choice is twofold.
First of all the large scale structure provides observational
constraints which are {\it a priori} independent of the CMB.  Second
of all, $f_{\rm NL}$ is not guaranteed to be scale independent, while
the LSS and CMB probe different scales. Indeed some inflationary
scenarios do predict large and scale-dependent $f_{\rm NL}$
\citep[see, e.g.,][]{chen2005,shandera2006}.

Departures from Gaussianity affect, among other things, the formation
and evolution of dark matter haloes of mass $M$.  One way of
quantifying the effect is to look at the halo mass function which we
can express as the product of the Gaussian mass function, $n_G(M,z)$,
times a NG correction factor, $F_{NG}(M,z,\fnl)$:
\beq\label{massfunction}
n(M,z,\fnl)=n_G(M,z)\;F_{NG}(M,z,\fnl)\ .
\eeq
In the extended Press-Schecter scenario, the NG factor can be written
as \citep[hereafter MVJ]{matarrese2000}
\bea\label{eq:rMVJ}
F_{NG}(M,z,\fnl)\simeq
\frac{1}{6}\frac{\delta_c^2(z_c)}{\delta_*(z_c)}\frac{d S_{3,M}}{d\ln \sigma_M}
+\frac{\delta_*(z_c)}{\delta_c(z_c)}\ ,
\eea
where $\delta_*(z_c)=\delta_c(z_c)\sqrt{1-S_{3,M}\delta_c(z_c)/3}$,
$\sigma_M^2$ is the mass variance at the mass scale $M$ (linearly
extrapolated to $z=0$), and $\delta_c(z_c)=\Delta_c/D_{+}(z_c)$ with
$D_{+}$ the growing mode of linear density fluctuations and $\Delta_c$
the linear extrapolation of the overdensity for spherical
collapse. The quantity $S_{3,M}$ represents the normalized skewness of
the primordial density field on scale $M$ (linearly extrapolated to
$z=0$), namely 
$S_{3,M}\equiv\frac{\langle\delta_M^3\rangle}{\sigma_M^4}\propto -
\fnl$ [see, e.g., eqs.(43-45) in MVJ, in which notation $\fnl \to
-\varepsilon$ in Model B]; $\delta_M$ represents the density
fluctuation smoothed on the mass scale $M$.  The redshift dependence is
only through $z_c$, the collapse redshift, which, in the extended
Press-Schechter approach, coincides with the considered epoch,
i.e. $z_c=z$.

A more robust statistics, that we will use throughout this paper, is
the ratio $R_{NG}(M,z,\fnl)$ of the NG cumulative mass function
$N_{\rm NG}(>M,z,\fnl)$ to the corresponding Gaussian one.  Since in
the high-mass tail $N(>M,z) \sim n(M,z) \times M$, we can approximate
$R_{NG}(M,z,\fnl)\simeq F_{NG}(M,z,\fnl)$ \citep{verde2001}.  We
notice that, being defined as a ratio, $R_{NG}$ is almost independent
of the explicit form assumed for $n_G(M,z)$.  We found, however, that
the mass function in our simulation with Gaussian initial conditions
is well fitted by the \cite{sheth1999} relation.

An alternative approximation for $R_{NG}$ can obtained starting from
eq.(62) of MVJ: $R_{NG}\simeq P_{NG}(>\delta_c|z,M)/
P_{G}(>\delta_c|z,M)$, where
\bea
P_{NG}(>\delta_c|z,M)\simeq & \frac{1}{2}-\frac{1}{\pi}\int_0^\infty
\frac{d\lambda}{\lambda} \exp\left( -\frac{\lambda^2 \sigma_M^2}{2}\right)
\nonumber \\
&\times \sin \left( \lambda \delta_c +\frac{\lambda^3 \langle 
\delta_M^3\rangle}
{6}\right)\ ,
\label{eq:eq62}
\eea
that reduces to the Gaussian case $P_{G}(>\delta_c|z,M)$ when $\langle
\delta_M^3\rangle=0$.  As we will see, this formula provides a better
fit for large positive values of $f_{\rm NL}$, while the previous one
should be preferred when $f_{\rm NL}$ is large and negative.

\section{Numerical simulations} \label{sect:simul}

We have used the GADGET-2 numerical code \citep{springel2005} to
perform two different sets of N-body simulations using collisionless,
dark matter particles only.

The first set consists of 7 different simulations characterized by
different values of $\fnl$ but the same $\Lambda$CDM model with mass
density parameter $\Omega_{m}=0.3$, baryon density $\Omega_{b}=0.04$,
Hubble parameter $h= H_{0}/(100 \ {\rm km} \ {\rm s}^{-1} \ {\rm
Mpc}^{-1})=0.7$, primordial power law index $n=1$ and $\sigma_8=0.9$,
in agreement with the WMAP first-year data \citep{spergel2003}.  In
all experiments we have loaded a computational box of
$500^{3}$(Mpc/h)$^{3}$ with $800^{3}$ dark matter particles, each one
with a mass of $m=2.033 \times 10^{10}$ M$_{\odot}$ h$^{-1}$. The
force was computed using a softening length $\epsilon_{l}=12.5$
h$^{-1}$ kpc.  To set the initial conditions we have used the same
initial Gaussian gravitational potential $\Phi_{\rm L}$ characterized
by a power-law spectrum $P(k) \propto k^{-3}$ which was then
inverse-Fourier transformed to get the NG term $f_{\rm{NL}}
\left(\Phi_{\rm L}^2 - \langle\Phi_{\rm L}^2\rangle
\right)$ in real space, and then transformed back to $k$-space, 
to account for the CDM matter transfer function. This procedure has
been adapted from the original one of \citep{moscardini1991} and
guarantees that all the $\fnl$ models have the same linear power
spectrum as the Gaussian case, as we checked in the different
realizations.  The gravitational potential was then used to displace
particles according to the Zel'dovich approximation, starting from a
glass-like distribution. Besides the Gaussian case ($\fnl=0$), we have
explored 6 different NG scenarios characterized by $\fnl=\pm 100 \;,
\pm 500 \; {\rm and}\; \pm 1000$.
 
The second set of simulations is designed to match those of 
\citep{kang2007}. Therefore, we have used smaller boxes of  
$300^{3}$(Mpc/h)$^{3}$ with $128^{3}$ particles and a different choice
of cosmological parameters ($\Omega_{m}=0.233$,
$\Omega_{\Lambda}=0.762$, $H_{0}=73 \ {\rm km} \ {\rm s}^{-1} \ {\rm
Mpc}^{-1}$ and $\sigma_8=0.74$), in agreement with the WMAP three-year
data \citep{spergel2007}.  We have used the same procedure to generate
the primordial NG, in order to explore the cases of
$\fnl=(-58,0,134)$, investigated by \citep{kang2007}.

\section{The Halo Mass Function} \label{sect:halomass}

To extract the dark matter haloes in the simulations we have used two
standard algorithms: the \emph{friends-of-friends} and the
\emph{spherical overdensity} methods.  The haloes, identified
using a linking length of 0.16 times the mean interparticle distance,
contain at least 32 particles.

\begin{figure*} 
\psfig{figure=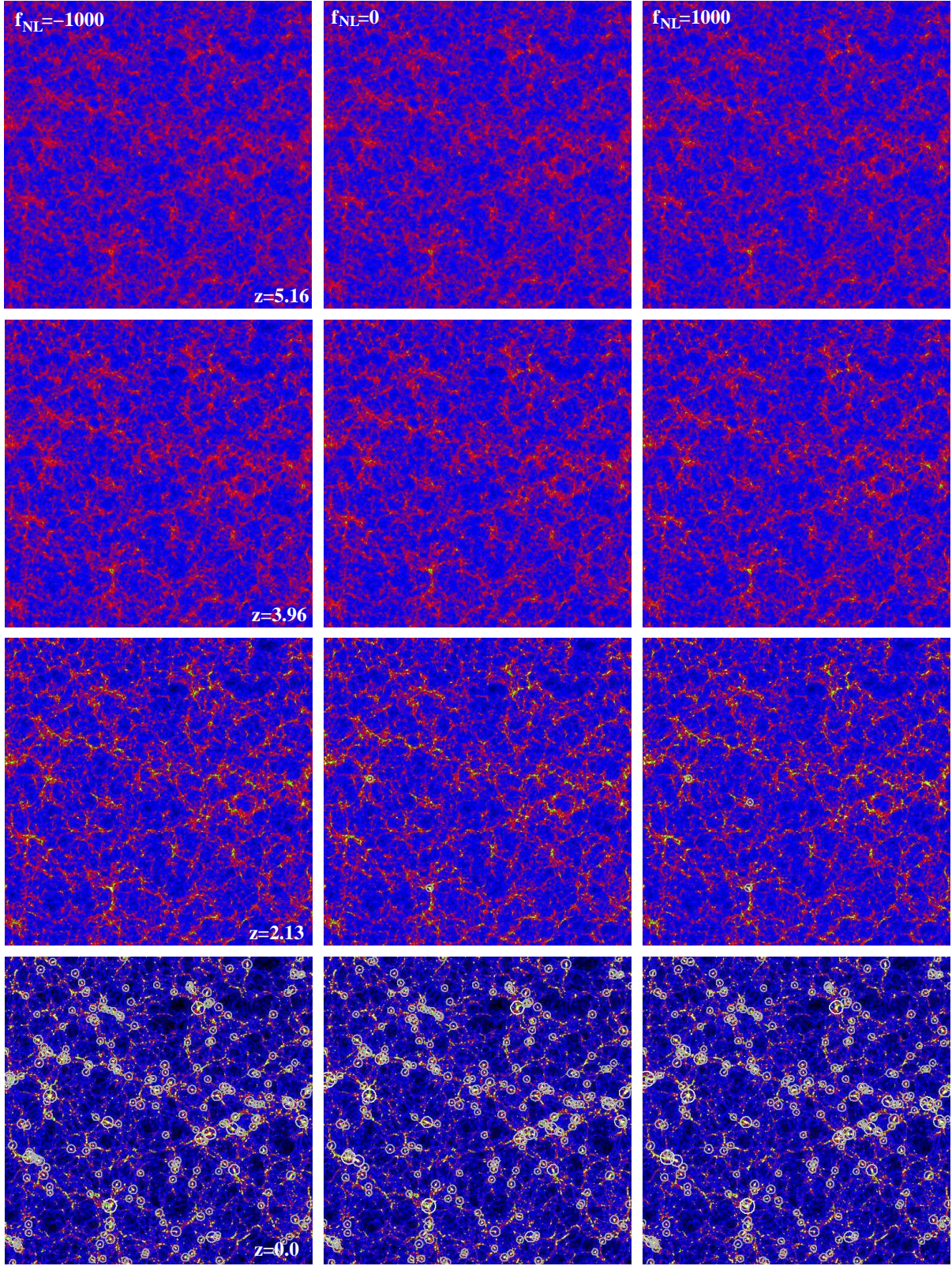,width=0.9\textwidth}
\caption{Mass density distribution and halo positions in a slice cut across
the simulation box. The color-coded contours indicate different
density levels ranging from dark (deep blue) underdense regions to
bright (yellow) high density peaks.  The halo positions are indicated
by open circles with size proportional to their masses.  Left panels:
NG model with $\fnl=-1000$. Central panels: Gaussian model.  Right
panels NG model with $\fnl=+1000$. The mass and halo distributions are
shown at various epochs, characterized by increasing redshifts (from
bottom to top), as indicated in the panels.}
\label{fig:slices} 
\end{figure*}

In Fig.~\ref{fig:slices} we show the spatial positions of the dark
haloes (circles) in the first set of simulations, superimposed on the
mass density field (colour-coded contour plots), in a redshift slice
of thickness 25.6 Mpc/h cut across the computational box.  The maps
are shown at four different redshifts, indicated in the plots, to
follow their relative evolution (from top to bottom). The panels in
the central column refer to the standard Gaussian case.  The two
extreme NG cases of $\fnl=-1000$ and $\fnl=+1000$ are illustrated in
the panels in the left and right columns, respectively. The mass
density fields look very similar in the Gaussian and NG cases at all
redshifts. Deviations from Gaussianity become apparent only when
considering, at a given epoch, the number and distribution of the
top-massive, virialized objects.  This is particularly evident in the
$z=2.13$ snapshot, in which the first-forming cluster-sized haloes can
only be seen in the $\fnl=+1000$ and $\fnl=0$ scenarios, in which
cosmic structures are indeed expected to form earlier with respect to
the $\fnl=-1000$ case.
 
\begin{figure*} 
\psfig{figure=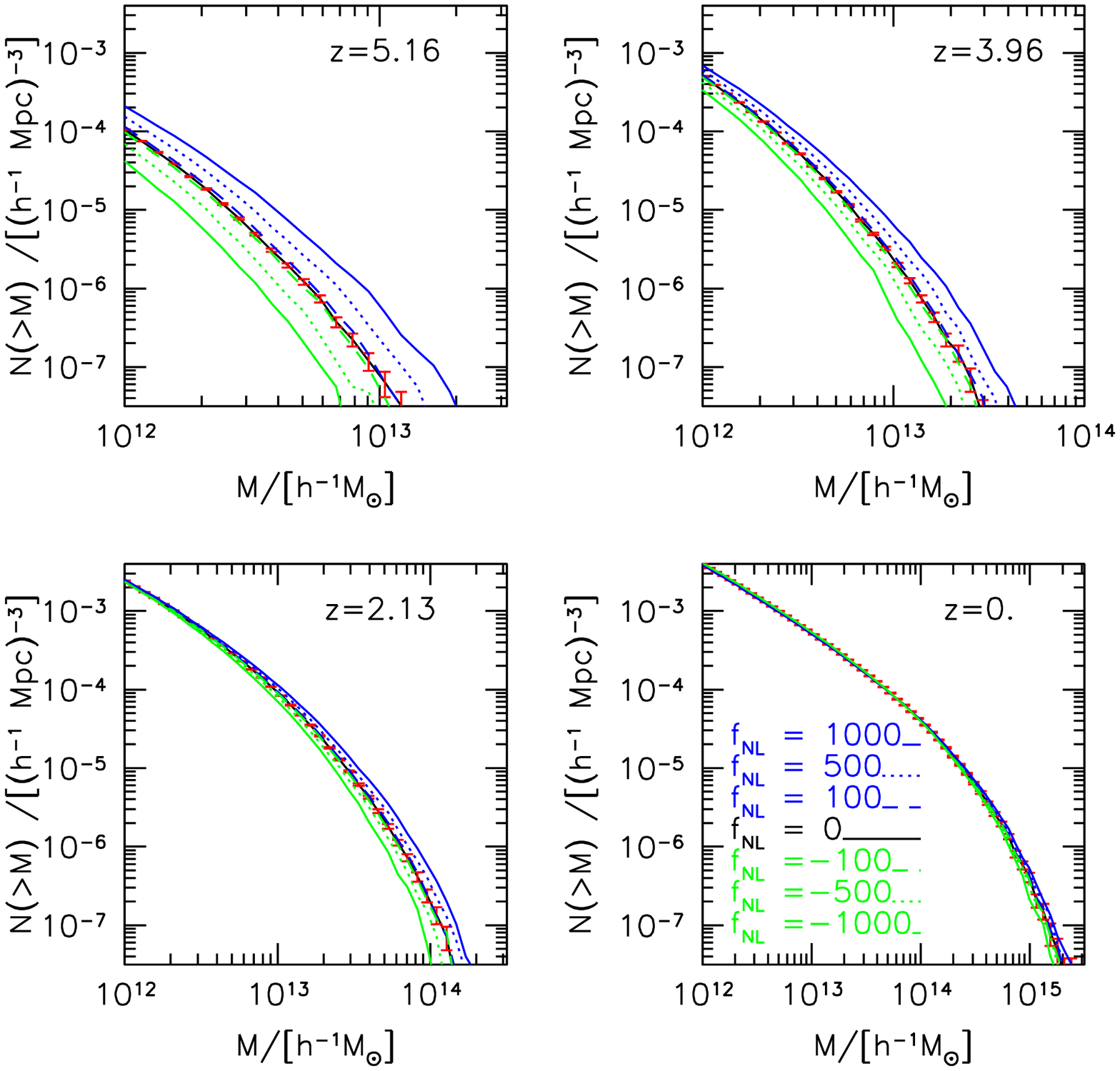,width=0.8\textwidth} 
\caption{Cumulative mass function of the haloes in the N-body experiments at 
the same redshifts as in Fig.~\ref{fig:slices}. The thick solid curve
represents the halo mass function in the Gaussian model.  Thin solid
line: $\fnl=\pm 1000$.  Dotted line $\fnl=\pm 500$. Dashed line
$\fnl=\pm 100$. The mass function of NG models with positive (negative)
$\fnl$ lies above (below) the Gaussian one.  Poisson errors are shown
for clarity only for the Gaussian model. }
\label{fig:cumul} 
\end{figure*}

The visual impression is corroborated by the plots of
Fig.~\ref{fig:cumul} in which we show the cumulative halo mass
function of all models in the first set of simulations, considered at
the same redshifts as Fig.~\ref{fig:slices}.  The number density of
massive objects increases with $\fnl$, as expected, and the
differences between models increases with the redshift, confirming
that the occurrence of massive objects at early epochs provides an
important observational test for NG models.

\begin{figure*} 
\psfig{figure=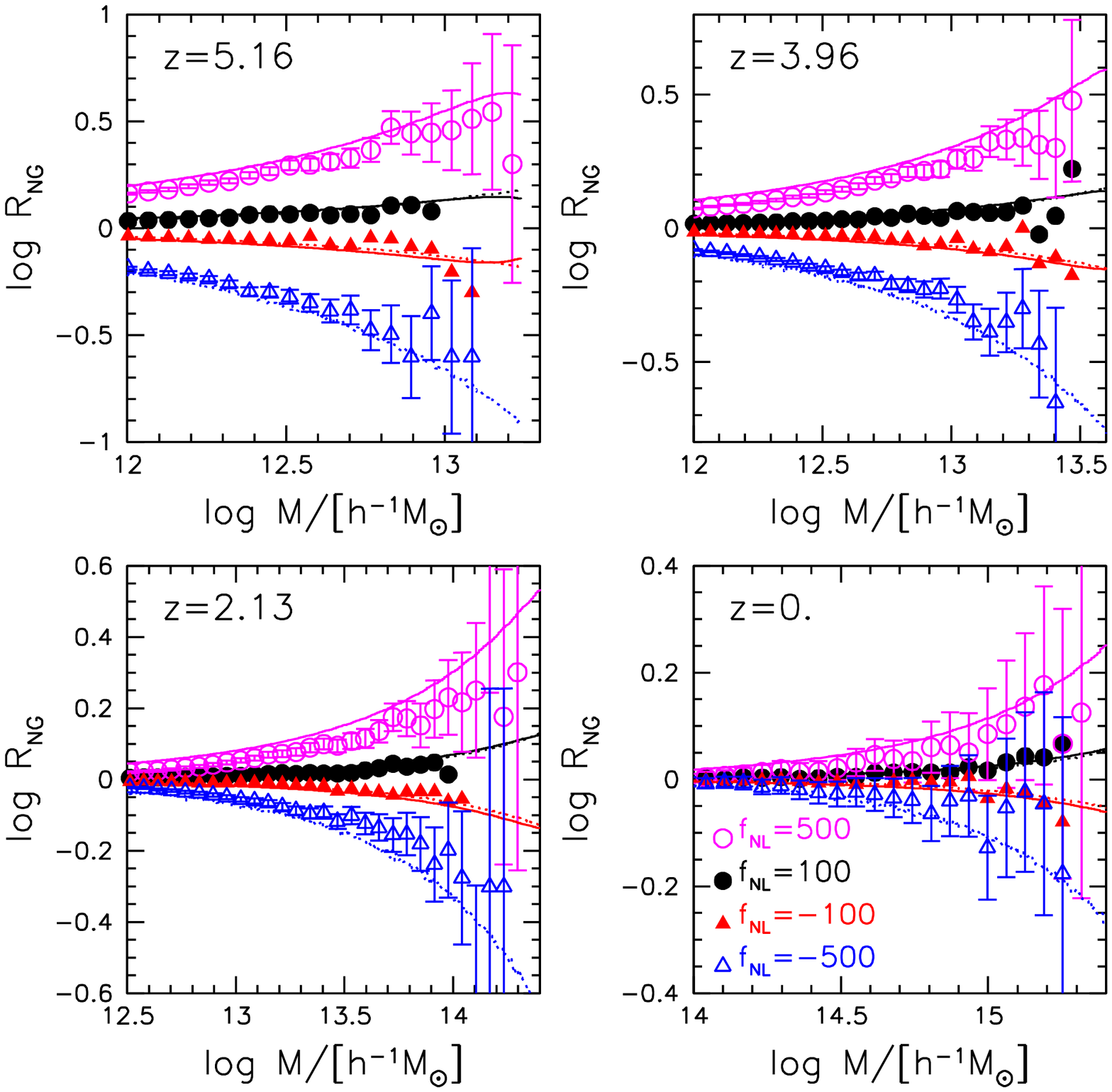,width=0.8\textwidth} 
\caption{Logarithm of the ratio of the halo cumulative
mass functions $R_{NG}$ as a function of the mass is shown in the
different panels at the same redshifts as in Fig.~\ref{fig:slices}.
Circles and triangles refer to positive and negative values for
$\fnl$; open and filled symbols refer to $\fnl=\pm 500$ and $\fnl=\pm
100$, respectively.  Theoretical predictions obtained starting from
eqs.~(\ref{eq:rMVJ}) and (\ref{eq:eq62}) are shown by dotted and solid
lines, respectively.  Poisson errors are shown for clarity only for
the cases $\fnl=\pm 500$.  }
\label{fig:ratio} 
\end{figure*}

In Fig.~\ref{fig:ratio} we show the logarithm of the ratio of the
cumulative halo mass functions of NG and Gaussian models,
$R_{NG}(M,z,\fnl)$, for our N-body simulations (symbols) compared to
the theoretical predictions of MVJ (solid curves).  Filled circles and
triangles, that refer to $\fnl=+100$ and $-100$, are compared to
theoretical predictions of both eqs.~(\ref{eq:rMVJ}) and
(\ref{eq:eq62}) (solid and dotted lines, respectively).  The agreement
between model and simulation is striking. Apparent deviations are well
within the Poisson errors that we do not show to avoid overcrowding.
Models and simulations are still in agreement, within the errors, for
$|\fnl|=500$ (open circles and triangles) and $|\fnl|=1000$ (not shown
in the figure) and out to $z\sim 5$, i.e.  when the validity of the
MVJ approximated expressions starts breaking down: we only notice that
the analytic model tends to slightly overpredict the difference
between NG and Gaussian cases.  In particular, we have found that
eq.~(\ref{eq:eq62}) is a better approximation to large, positive
$\fnl$, while eq.~(\ref{eq:rMVJ}) is to be preferred when $\fnl$ is
negative.

This result is at variance with that of \cite{kang2007} in which the
disagreement with the MVJ predictions is already significant at the
more moderate values of $\fnl=-58$ and $\fnl=+134$, as it is clearly
illustrated in the bottom panels of their Fig.~2.  To investigate
whether the mismatch is genuine or is to be ascribed to numerical
effects we have repeated the same analysis using the second set of
simulations. We still found an excellent agreement between numerical
experiments and theoretical expectations, which adds to the robustness
of our results.

\section{Discussion and Conclusions} \label{sect:discussion}

The numerical experiments performed in this work show the validity of
the analytic model proposed by \cite{matarrese2000} for the halo mass
function in NG scenarios. The model fits the results of our numerical
experiments over a large range of $\fnl$ and out to high redshifts
despite the fact that non-Gaussianity is treated as perturbation of
the underlying Gaussian model. The good match with the numerical
experiments is, however, not surprising, since deviations from
Gaussianity in the primordial gravitational potential are within the
percent level, despite the large values used for $\fnl$.  This result
is in disagreement with that of \cite{kang2007} who found a number
density of haloes larger (smaller) than ours for NG models with the
same positive (negative) $\fnl$ values.  The mismatch can be ascribed
neither to resolution effects nor to the background cosmology adopted
in the numerical experiments and, perhaps, should be traced back to a
different setup in the NG initial conditions.

Our result has two important consequences.  The good news is that we
can use an analytical model to provide accurate predictions for the
mean halo abundance in NG scenarios spanning a range of $\fnl$ well
beyond the current WMAP constraints, without relying on time-consuming
numerical experiments affected by shot noise and cosmic variance.  The
bad news is that deviations from the Gaussian model are more modest
than in the \cite{kang2007} experiments, which makes more difficult to
spot non-Gaussianity from the analysis of the LSS.  This, however,
might not be a problem when considering the observational constraints
that will be provided by next-generation surveys.  Indeed, current
hints of non-Gaussianity from large scale structures are rather weak
as they are provided by the excess power in the distribution of 2dF
galaxies at $z \sim 0$ \citep{norberg2002,baugh2004}, for which it is
difficult to disentangle primordial non-Gaussianity from galaxy bias,
and by the possible presence of protoclusters at $z\sim 4$, as traced
by radio galaxies surrounded by Ly-$\alpha$ emitters and Ly-break
galaxies \citep{miley2004}, for which the measured velocity dispersion
can be compared to theoretical predictions only including a realistic
model for the velocity bias between galaxies and dark matter.
However, the observational situation is going to improve dramatically,
in many respects.  Next-generation cluster surveys in X-ray, microwave
and millimeter bands like the upcoming eROSITA X-ray cluster survey
\citep{predehl2006}, the South Pole Telescope survey \citep{ruhl2004},
the Dark Energy Survey \citep{abbott2005} and the Atacama Cosmology
Telescope survey \citep{kosowsky2006} should allow us to detect
clusters out to high redshift.  The abundance of these objects is,
however, more sensitive to the presence of a dark energy component
than to that of a primordial non-Gaussianity, at least for $|\fnl| \le
100$.  Indeed, as pointed out by \cite{sefusatti2007}, dark energy
constraints from these surveys will not be substantially affected by
primordial non-Gaussianity, as long as deviations from the Gaussian
model do not dramatically exceed current WMAP constraints. However,
these surveys could provide a fundamental cross-check on any detection
of non-Gaussianity from CMB, especially from Planck since the physical
scales probed by clusters differ from that of Planck by only a factor
of two, allowing us to detect a possible scale dependence of
non-Gaussian features.  More effective constraints on both dark energy
and non-Gaussianity based on statistics of rare objects is likely to
be provided by the imaging of the mass distribution from gravitational
lensing of high-redshift 21 cm absorption/emission signal that, if
observed with a resolution of a few arcsec, will allow us to detect
haloes more massive than the Milky Way back to $z\sim10$
\citep{metcalf2006}.  

Tight constraints on primordial non-Gaussianity can be set by
measuring the high-order statistics of the LSS. In particular, the
bispectrum analysis of the galaxy distribution at moderate $z$ in the
next-generation redshift surveys such as HETDEX \citep{hill2004} and
WFMOS2 \citep{glazebrook2005} will efficiently disentangle non-linear
biasing from primordial non-Gaussianity \citep{sefusatti2007b}.

In this work we have demonstrated the validity of the
\cite{matarrese2000} approach to detect non-Gaussianity from the
statistics of rare objects. In a future paper we will extend our
analysis by considering the Probability Distribution Function of
density fluctuations, that is easily obtained from our numerical
simulations and that in principle can be determined by measuring the
21 cm line, either in emission or in absorption, before or after the
epoch of reionization \citep{furlanetto2006}.

\section*{acknowledgments}
Computations have been performed by using 128 processors on the
IBM-SP5 at CINECA (Consorzio Interuniversitario del Nord-Est per il
Calcolo Automatico), Bologna, with CPU time assigned under an
INAF-CINECA grant and on the IBM-SP4 machine at the ``Rechenzentrum
der Max-Planck-Gesellschaft'' at the Max-Planck Institut fuer
Plasmaphysik with CPU time assigned to the MPA.  We acknowledge
financial contribution from contracts ASI-INAF I/023/05/0, ASI-INAF
I/088/06/0 and INFN PD51. We thank Licia Verde for useful discussions,
Shude Mao for his critical reading of the manuscript, and Claudio
Gheller for his assistance.

\label{lastpage}
\end{document}